\documentclass{aa}
\usepackage{graphics,epsfig,amsmath,amssymb,amstext,txfonts}

\def\d {\mathrm{d}}

\def\la{\mathrel{\mathchoice {\vcenter{\offinterlineskip\halign{\hfil
$\displaystyle##$\hfil\cr<\cr\sim\cr}}}
{\vcenter{\offinterlineskip\halign{\hfil$\textstyle##$\hfil\cr
<\cr\sim\cr}}}
{\vcenter{\offinterlineskip\halign{\hfil$\scriptstyle##$\hfil\cr
<\cr\sim\cr}}}
{\vcenter{\offinterlineskip\halign{\hfil$\scriptscriptstyle##$\hfil\cr
<\cr\sim\cr}}}}}

\def\ga{\mathrel{\mathchoice {\vcenter{\offinterlineskip\halign{\hfil
$\displaystyle##$\hfil\cr>\cr\sim\cr}}}
{\vcenter{\offinterlineskip\halign{\hfil$\textstyle##$\hfil\cr
>\cr\sim\cr}}}
{\vcenter{\offinterlineskip\halign{\hfil$\scriptstyle##$\hfil\cr
>\cr\sim\cr}}}
{\vcenter{\offinterlineskip\halign{\hfil$\scriptscriptstyle##$\hfil\cr
>\cr\sim\cr}}}}}

\begin{document}

\title{Generalised 3D-reconstruction method of a dipole anisotropy in cosmic-ray distributions}

\author{J. Aublin \& E. Parizot}

\offprints{parizot@ipno.in2p3.fr}

\institute{Institut de Physique Nucl\'eaire d'Orsay, IN2P3-CNRS/Universit\'e Paris-Sud, 91406 Orsay Cedex, France.}

\date{[Accepted for publication in A\&A: 07/04/2005]}

\abstract{We develop a method to study the anisotropy of a cosmic-ray angular distribution, using both the right ascension and the declination of the arrival directions. It generalises the full-sky coverage method of Sommers (2001) to partial-sky coverage experiments. When the angular distribution consists of a dipolar modulation of an otherwise isotropic flux, the method allows one to reconstruct the dipole amplitude and the dipole orientation in 3D space. We analyse in detail the statistical properties of the method, introducing the concept of reconstruction power, and show that it is generally more powerful than the standard Rayleigh analysis in right ascension. We clarify the link between the traditionally-used first harmonic amplitude and the true, physical dipole amplitude, and we investigate the variation of the reconstruction powers as a function of the dipole orientation. We illustrate the method by computing the amplitude and angular reconstruction powers of the Pierre Auger Observatory, with the Southern site alone and with both Southern and Northern sites. In this particular case, we find that the building of a similar site in the Northern hemisphere would decrease the time needed for the method to reveal a significant departure from an isotropic cosmic-ray distribution by a factor of about eight.
}

\maketitle

\section{Introduction}
\label{sec:introduction}

Although cosmic-rays (CRs) have been known for almost one century, their origin remains uncertain, mostly because their trajectories are bent by Galactic magnetic fields and they do not individually point back to their sources. Moreover, since these fields are chaotic on scales ranging at least from $10^{8}\,\mathrm{cm}$ to $10^{20}\,\mathrm{cm}$ (Armstrong et al. 1995), the transport of CRs is diffusive up to high energies, which tends to make their angular distribution isotropic. Therefore, even collectively, the CR arrival directions hold virtually no information about the source distribution in space.

However, as the energy of the CRs increases, anisotropies can appear either because the diffusive approximation does not hold anymore, or because the diffusion coefficient becomes large enough to reveal intrinsic inhomogeneities in the source distribution. Specifically, even if the diffusive regime holds, the density of CR sources in the Galaxy is believed to be larger in the inner regions than in the outer ones, and this can cause a slightly higher CR flux coming from the Galactic centre (GC) than from the anti-center. Likewise, the global CR streaming away from the Galactic plane (towards the halo) can be a source of measurable anisotropy. However, the detailed angular distribution of CRs is quite hard to predict, even if we assume a definite source distribution, because it also depends on the propagation conditions, which are related to both large scale and small scale magnetic field configurations, and on the position of the Earth relative to major magnetic structures, such as the local Galactic arm. The presence of numerous nearby superbubbles, which can break out the Galactic disk and produce chimneys (e.g. de Avillez \& Berry 2001) along which CRs diffuse more easily into the halo, can also be responsible for specific CR anisotropies.

Under the simplest assumptions (isotropic diffusion, homogeneous distribution of sources in the Galactic disk), the off-centered position of the Earth in the Galaxy (radially and vertically) leads to a dominant dipole anisotropy with an amplitude $\alpha$ proportional to the energy-dependent diffusion coefficient, $D(E)$ (e.g. Jones et al. 2001). At higher energy, it is also expected that the influence of local structures will become less important. A dipole anisotropy, although with a small amplitude, can also arise because of the relative motion of the solar system with respect to the interstellar plasma carrying the diffusing centers. At ultra-high-energy (UHE), a dipole can also appear if CRs propagate in straight lines from exotic sources distributed all over the Galactic halo, e.g. related to the dark matter (e.g. Berezinsky 2000), or if a roughly diffusive propagation settles between a dominant source (e.g. the Virgo cluster) and our Galaxy. Higher order multipoles can also arise naturally in some models. For instance, a dominant quadrupole in the UHECR angular distribution could result from an equatorial excess of sources in supergalactic coordinates (Sommers 2001).

From a general point of view, the characterisation of the CR anisotropy provides useful information to constrain the GCR diffusion models, notably the effective diffusion coefficients, related to the magnetic field structure. Indeed, the level of CR anisotropy depends on the diffusion coefficient: in a simple model where CR sources are homogeneously distributed in a disk of thickness $2h$ and the CRs are confined in a halo of height $H$, the anisotropy at a distance $z$ above the Galactic plane ($z<h$) is estimated as $\delta\simeq 3D/cH\times z/h$ (Ptuskin 1997). Anisotropy measurements at various energies can thus provide crucial information about the energy dependence of the diffusion coefficient. This information is particularly important to constrain the GCR source spectrum, since it sets the relation between the source power-law index and the observed one, through the energy dependent confinement of CRs in the Galaxy.

In Sect.~\ref{sec:dipoleReconstruction}, we present a method to derive the parameters of an assumed dipole anisotropy in the CR angular distribution, following Sommers (2001) in using both the right ascension, RA, and the declination, $\delta$, of a discrete set of CRs. This is in contrast with the standard ``Rayleigh analysis'' (i.e. harmonic analysis on the circle) used in cosmic-ray physics, which makes use of the CR distribution in right ascension only (Linsley 1975). The method exposed below consists essentially of a generalization of Sommers' method (i.e. harmonic analysis on the sphere, Sommers 2001) to limited sky coverage. In Sect.~\ref{sec:accuracy}, we study quantitatively the statistical power of the method for the reconstruction of the dipole amplitude and orientation, in the 3-dimensional space (two angular dimensions). We then apply this study to the case of the Pierre Auger Observatory (PAO, e.g. Auger Collaboration 2004), which will soon obtain the largest data set in the crucial energy range between $10^{18}$ and $5\,10^{18}$~eV, i.e. at the end of the GCR component, before the transition to extragalactic CRs. We then compare the accuracy of the Rayleigh analysis with that of our 3D method. We also discuss the increase in the reconstruction power which will be provided by the second site of the PAO, to be installed in the Northern hemisphere.

\section{Dipole reconstruction procedure}
\label{sec:dipoleReconstruction}

\subsection{Anisotropy in right ascension: the Rayleigh analysis}
\label{sec:RayleighMethod}

The standard way to estimate the anisotropy of the CR distribution at a given energy consists in performing a Fourier analysis of the CR arrival directions and computing the first-harmonic amplitude in right ascension, $r_{\mathrm{1h}}$. This is obtained from the sums of the sine and cosine of the right ascension of each of the $N$ events, which is also the azimuthal angle, $\varphi$, in equatorial spherical coordinates:
\begin{equation}
a = \frac{2}{N}\sum_{i=1}^N \cos\varphi_{i}\quad\mathrm{and}\quad b = \frac{2}{N}\sum_{i=1}^N \sin\varphi_{i},
\label{eq:aAndB}
\end{equation}
as
\begin{equation}
r_{\mathrm{1h}} = \sqrt{a^2 + b^2}.
\label{eq:r1h}
\end{equation}
The direction of the anisotropy is also obtained from the phase of the first harmonic in right ascension:
\begin{equation}
\psi_{\mathrm{1h}} = \arctan{\frac{b}{a}}.
\label{eq:psi1h}
\end{equation}

The reconstructed values of the first harmonic amplitude and phase converge towards the actual ones when the number of CRs tends to infinity. The distribution of reconstructed values for limited data sets has been analysed formally by Linsley (1975). For sufficiently large numbers of events, $N\gg 4/r_{\mathrm{1h}}^2$, it is well described by a Gaussian distribution with widths $\sigma_{r} = \sqrt{2/N}$ and $\sigma_{\psi} = \sqrt{2/Nr_{\mathrm{1h}}^2}$. For the purpose of comparison with the 2D approach described below, we can reformulate this result by defining the \emph{significance} of the measurement of a given anisotropy amplitude, $r_{\mathrm{1h}}$, as the ``number of sigmas'':
\begin{equation}
n_{\sigma} = \frac{r_{\mathrm{1h}}}{\sigma_{r}} = \frac{1}{\sqrt{2}}\,r_{\mathrm{1h}}\,\sqrt{N}.
\label{eq:nSigmaR1h}
\end{equation}

With the above method, the direction of the flux excess is only known by its RA coordinate, since no information about the declination is available. This is an obvious consequence of the initial choice of marginalising with respect to this variable, which is usually made for practical reasons. The acceptance of a CR experiment usually depends much more on declination than on right ascension, because all the points of the sky at the same declination are observed in roughly the same conditions as the Earth rotates around its poles. In the case of experiments working around 1--100~TeV, the CR anisotropy is so small (see Sect.~\ref{sec:introduction}) that even a limited uncertainty of the order of 1\% in the declination dependence of the acceptance would completely blur the results.

At higher energies, however, the anisotropy is expected to be higher, and the determination of its direction, in right ascension \emph{and} declination, appears both possible and desirable. In the next subsection, we recall the method proposed by Sommers (2001) in the case of a full sky coverage experiment, and then we give a generalization for partial-sky coverage.

\subsection{Complete dipole reconstruction with full sky coverage}

Let us start by writing the assumed shape of the CR angular distribution:
\begin{equation}
\Phi(\mathbf{u}) = \frac{\Phi_{0}}{4\pi}(1 + \alpha \mathbf{D}\cdot\mathbf{u}),
\label{eq:CRDistribWithDipole}
\end{equation}
where the differential flux, $\Phi$, in direction $\mathbf{u}$, consists of an isotropic part, $\Phi_{0}/4\pi$, modulated by a dipolar component, in $\cos(\widehat{\mathbf{u},\mathbf{D}})$. Here, $\mathbf{D}$ is the unit vector pointing in the direction of the dipole, and $\alpha$ is the dipole amplitude, relative to the monopole: $0\le \alpha\le 1$. This can be seen as the first order development in spherical harmonics of the CR angular distribution. In the following, we shall always assume that the higher order terms (quadrupole, etc.) are negligible, so that the flux which we want to reconstruct is of the form of Eq.~(\ref{eq:CRDistribWithDipole}). Specifically, we want to derive $\alpha$ and $\mathbf{D}$ from the data, i.e. three parameters: one amplitude, and two angles.

The idea is to compute some integral quantities and to identify their theoretical values with the discrete versions provided by discrete sums over experimental data. In order to reconstruct the three parameters of a dipole, we need three quantities and an additional one corresponding to the global flux normalization. This is provided by the following moments of order zero and one:
\begin{equation}
I_{0} = \int \Phi(\mathbf{u})\d\Omega \quad\mathrm{and}\quad \mathbf{I} = \int \mathbf{u}\Phi(\mathbf{u})\d\Omega.
\label{eq:0thAnd1stMoments}
\end{equation}

It is straightforward to obtain these integrals over the whole sky when $\Phi(\mathbf{u})$ is given by Eq.~(\ref{eq:CRDistribWithDipole}):
\begin{equation}
I_{0} = \Phi_{0} \quad\mathrm{and}\quad \mathbf{I} = \frac{1}{3}\Phi_{0}\times \alpha \,\mathbf{D}.
\label{eq:I0AndIFullSky}
\end{equation}

The discrete version of these integrals, $S_{0}$ and $\mathbf{S}$, are obtained by dividing the sky into a series of pixels $\{(i,j)\}$, with solid angle $\delta\omega_{i,j}$, and changing the continuous integral into a sum over all the pixels:
\begin{equation}
\int f(\mathbf{u})\d\Omega \longrightarrow\sum_{(i,j)} f_{i,j}\,\delta\omega_{i,j}\,.
\label{eq:integralsToSums}
\end{equation}

Let $N_{i,j}$ be the number of CRs observed in the direction of the pixel $(i,j)$. Introducing the \emph{exposure} of a given direction on the sky, $\mathcal{E}(\mathbf{u})$ (in $\mathrm{m}^2\,\mathrm{s}$), relevant to a particular CR experiment, one can write the differential number of events observed in direction $\mathbf{u}$, within $\d\Omega$, as:
\begin{equation}
\frac{\d N}{\d\Omega}(\mathbf{u})\d\Omega = \Phi(\mathbf{u})\mathcal{E}(\mathbf{u})\d\Omega \longrightarrow N_{i,j} = \Phi_{i,j}\mathcal{E}_{i,j}\delta\omega_{i,j}\,.
\label{eq:nbOfEventsInPixelIJ}
\end{equation}

Transforming the integrals in Eq.~(\ref{eq:0thAnd1stMoments}) according to~(\ref{eq:integralsToSums}), and replacing $\delta\omega_{i,j}$ from Eq.~(\ref{eq:nbOfEventsInPixelIJ}), one obtains:
\begin{equation}
I_{0}\rightarrow S_{0} = \sum_{(i,j)}\frac{N_{i,j}}{\mathcal{E}_{i,j}} \quad\mathrm{and}\quad \mathbf{I}\rightarrow \mathbf{S} = \sum_{(i,j)}\frac{N_{i,j}\mathbf{u}_{i,j}}{\mathcal{E}_{i,j}}\,,
\label{eq:discreteSumsFullSky}
\end{equation}
where $\mathbf{u}_{i,j}$ is the unit vector in the direction of pixel $(i,j)$.

The last step consists in changing the sum over all directions (or pixels) into a sum over all events. This is done by noting that when one sums over all the events in the same pixel, $(i,j)$, one actually adds up the corresponding value $N_{i,j}$ times. Therefore, for any quantity $F$, if $F_{i,j}$ denotes its value in direction $(i,j)$, and $F_{k}$ is its value in the direction of event $k$, one can write: \begin{equation}
\sum_{(i,j)}N_{i,j}F_{i,j} = \sum_{k}F_{k},
\label{eq:pixelsToEvents}
\end{equation}
where the first sum is over all the pixels, and the second is over all the events.

Applying this to Eqs.~(\ref{eq:discreteSumsFullSky}), one finally gets the discrete versions of Eq.~(\ref{eq:0thAnd1stMoments}):
\begin{equation}
S_{0} = \sum_{k}\frac{1}{\mathcal{E}_{k}} \quad\mathrm{and}\quad \mathbf{S} = \sum_{k}\frac{\mathbf{u}_{k}}{\mathcal{E}_{k}}\,,
\label{eq:discreteSumsOverEventsFullSky}
\end{equation}
where the sums are over all the events and $\mathcal{E}_{k}$ is the exposure of the sky in the direction of event $k$, namely $\mathbf{u}_{k}$, as observed with the experiment under consideration.

The above four discrete sums, $S_{0}$ and $\mathbf{S}$, can be straightforwardly computed from the data (provided that the sky exposure is known, which simply derives from the detector's aperture). The derivation of the dipole parameters then follow directly from the identification with $I_{0}$ and $\mathbf{I}$, as given by Eqs.~(\ref{eq:I0AndIFullSky}). The dipole amplitude and directions are estimated (given the finite set of events available) as:
\begin{equation}
\alpha = 3\frac{||\mathbf{S}||}{S_{0}} \quad\mathrm{and}\quad \mathbf{D} = \frac{\mathbf{S}}{||\mathbf{S}||}.
\label{eq:identificationFullSky}
\end{equation}

Obviously, the above estimates are all the more accurate that the total number of events, $N$, is large (see Sect.~\ref{sec:accuracy}).

\subsection{Generalization to the case of partial sky coverage}

As noted by Sommers (2001), the discrete sums of Eq.~(\ref{eq:discreteSumsFullSky}) are meaningless in regions where the exposure is null. This prevents one from using the method in its current form with partial-sky coverage data sets. However, the above derivation makes it clear that a generalisation to such a case is possible, although slightly more complicated from the algebraic point of view. The simplicity of the method was due to the fact that the matrix giving $I_{0}$ and $\mathbf{I}$ as a function of the dipole parameters, $\alpha$ and $\mathbf{D}$, could be straightforwardly inverted to give $\alpha$ and $\mathbf{D}$ as a function of the integral moments, or their approximations $S_{0}$ and $\mathbf{S}$, as summed up in Eqs.~(\ref{eq:identificationFullSky}). Exactly the same procedure can be followed in the case of a limited sky coverage experiment.

Since the sums in Eqs.~(\ref{eq:identificationFullSky}) are over detected events, it is clear that the exposures, $\mathcal{E}_{k}$, in the denominator can never be zero, even though there are parts of the sky where the exposure is null. To identify $S_{0}$ and $\mathbf{S}$ with the corresponding moments, $I_{0}$ and $\mathbf{I}$, one simply needs to integrate the latter only over the part of the sky which is actually observed. Interestingly, the sky region covered by any terrestrial observatory (operating for a long enough period of time) will always be limited by lines of constant declination, i.e. independent of RA. This is an obvious consequence of the Earth's rotation around its axis and the Sun. In the following, we will thus assume that the part of the sky where the exposure is non-zero is contained between declination $\delta_{\mathrm{min}}$ and $\delta_{\mathrm{max}}$ (in equatorial coordinates), corresponding to spherical $\theta$ coordinates (as measured from the North Pole) between $\theta_{\mathrm{min}}$ and $\theta_{\mathrm{max}}$. Note that we do not require the exposure to be independent of RA (which is usually not the case, because of seasonal effects, e.g. related to temperature variations of the acceptance), but only that if a point on the celestial sphere with declination $\delta$ can be observed by the detector, then all the points at the same declination can also be observed (possibly with a different exposure), whatever their right ascension.

We thus rewrite the zeroth and first order moments of the CR intensity as:
\begin{equation}
\begin{split}
&I_{0} = \int_{\theta_{\mathrm{min}}}^{\theta_{\mathrm{max}}}\hspace{-5pt}\d\theta \,\sin\theta \int_{0}^{2\pi}\hspace{-5pt}\d\varphi \,\Phi(\mathbf{u}) \\
\mathrm{and}\quad &\mathbf{I} = \int_{\theta_{\mathrm{min}}}^{\theta_{\mathrm{max}}}\hspace{-5pt}\d\theta \,\sin\theta \int_{0}^{2\pi}\hspace{-5pt}\d\varphi \,\mathbf{u}\Phi(\mathbf{u}),
\end{split}
\label{eq:0thAnd1stMomentsLimitedSky}
\end{equation}
which can be integrated using Eq.~(\ref{eq:CRDistribWithDipole}) as:
\begin{equation}
\begin{split}
I_{0} &= \frac{\Phi_{0}d}{4}(2 + s \alpha D_{z})\\
I_{x} &= \frac{\Phi_{0}d}{4}(1 - \gamma)\alpha D_{x}\\
I_{y} &= \frac{\Phi_{0}d}{4}(1 - \gamma)\alpha D_{y}\\
I_{z} &= \frac{\Phi_{0}d}{4}(s + 2\gamma\alpha D_{z})\\
\end{split}
\label{eq:I0IxIyIz}
\end{equation}
where we have introduced the difference, sum and product
\begin{equation}
\begin{split}
d &= \cos\theta_{\mathrm{min}} - \cos\theta_{\mathrm{max}} \\
s &= \cos\theta_{\mathrm{min}} + \cos\theta_{\mathrm{max}} \\
p &= \cos\theta_{\mathrm{min}}\times\cos\theta_{\mathrm{max}}
\end{split}
\label{eq:dAndS}
\end{equation}
and defined
\begin{equation}
\gamma \equiv \frac{s^2-p}{3}.
\label{eq:gamma}
\end{equation}

Finally, Eqs.~(\ref{eq:I0IxIyIz}) can be inverted to obtain
\begin{equation}
\begin{split}
\alpha D_{x} &= \frac{I_{x}}{sI_{z} - 2\gamma I_{0}}\,\frac{\gamma - p}{\gamma - 1}\\
\alpha D_{y} &= \frac{I_{y}}{sI_{z} - 2\gamma I_{0}}\,\frac{\gamma - p}{\gamma - 1}\\
\alpha D_{z} &= \frac{sI_{0} - 2I_{z}}{sI_{z} - 2\gamma I_{0}}\\
\end{split}
\label{eq:alphaDxDyDz}
\end{equation}

As in the case of a full sky coverage, the dipole parameters can thus be evaluated straightforwardly from the discrete versions of $I_{0}$ and $\mathbf{I}$, which are still given by Eqs.~(\ref{eq:discreteSumsOverEventsFullSky}). The dipole direction is reconstructed as the direction of the vector on the right hand side of Eqs.~(\ref{eq:alphaDxDyDz}), while its amplitude is its norm.

For instance, the reconstructed dipole amplitude, $\alpha_{\mathrm{rec}}$, writes:
\begin{equation}
\alpha_{\mathrm{rec}}^2 = \frac{(\frac{\gamma - p}{\gamma-1})^2 (S_{x}^2 + S_{y}^2) + (sS_{0} - 2S_{z})^2}{(2\gamma S_{0} - sS_{z})^2},
\label{eq:alphaRec}
\end{equation}
which reduces to $\alpha_{\mathrm{rec}}^2 = 9(S_{x}^2 + S_{y}^2 + S_{z}^2)/S_{0}^2$, as expected, in the case of a full sky coverage experiment ($s=0$, $d=2$, $p=-1$, $\gamma=1/3$).

\begin{figure}[t]
\begin{center}
\epsfig{file=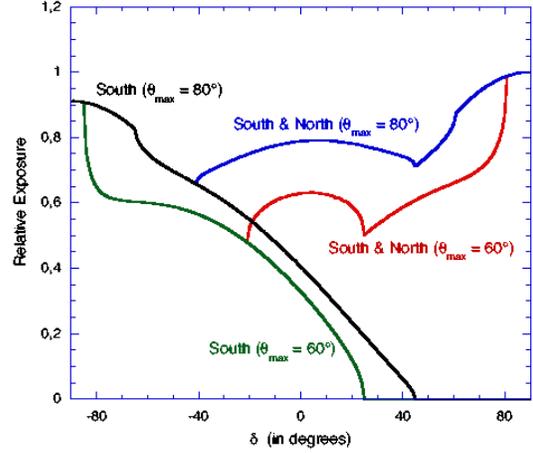,width=7cm}
\end{center}
\caption{Relative exposure of the PAO surface detector (either one or two sites) as a function of declination in equatorial coordinates, assuming efficient detection up to a maximum zenith angle of either 60$^\circ$ or 80$^\circ$.}
\label{fig:expo}
\end{figure}

An obvious advantage of this method with respect to the 2D Rayleigh method is that it entirely characterises the dipole vector, i.e. the direction of the anisotropy both in declination and right ascension. One may be worried, however, that this is at the expense of reconstruction accuracy, since it is essentially the same data set which is used to derive one more (angular) information. We show in the next section that this is not the case, and that the dipole reconstruction accuracy is in fact generally better with the method presented here.

\section{Reconstruction accuracy}
\label{sec:accuracy}

\subsection{Monte-Carlo technique}

\begin{figure*}[t]
\includegraphics[height=7.2cm]{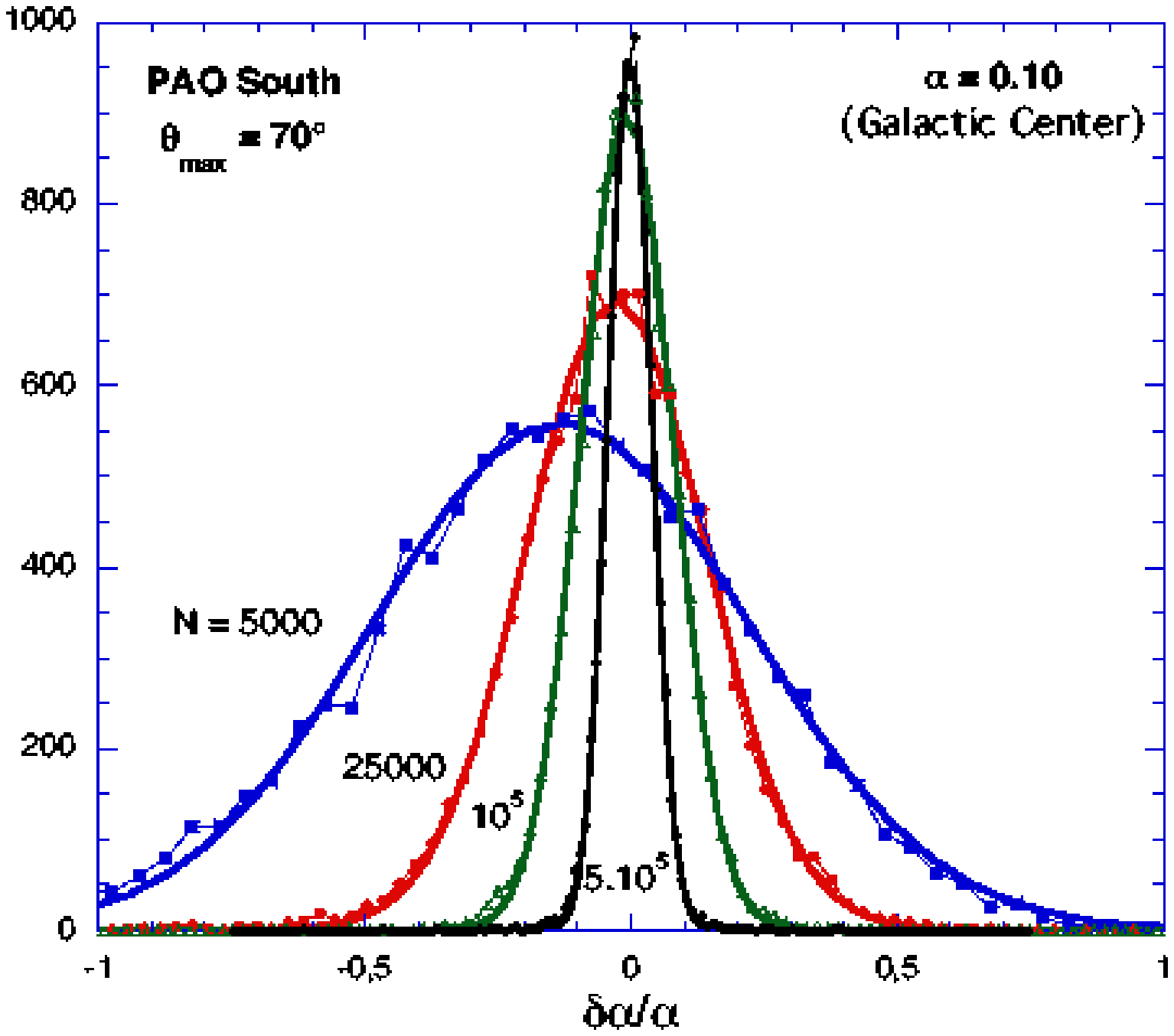}
\hfill
\includegraphics[height=7.2cm]{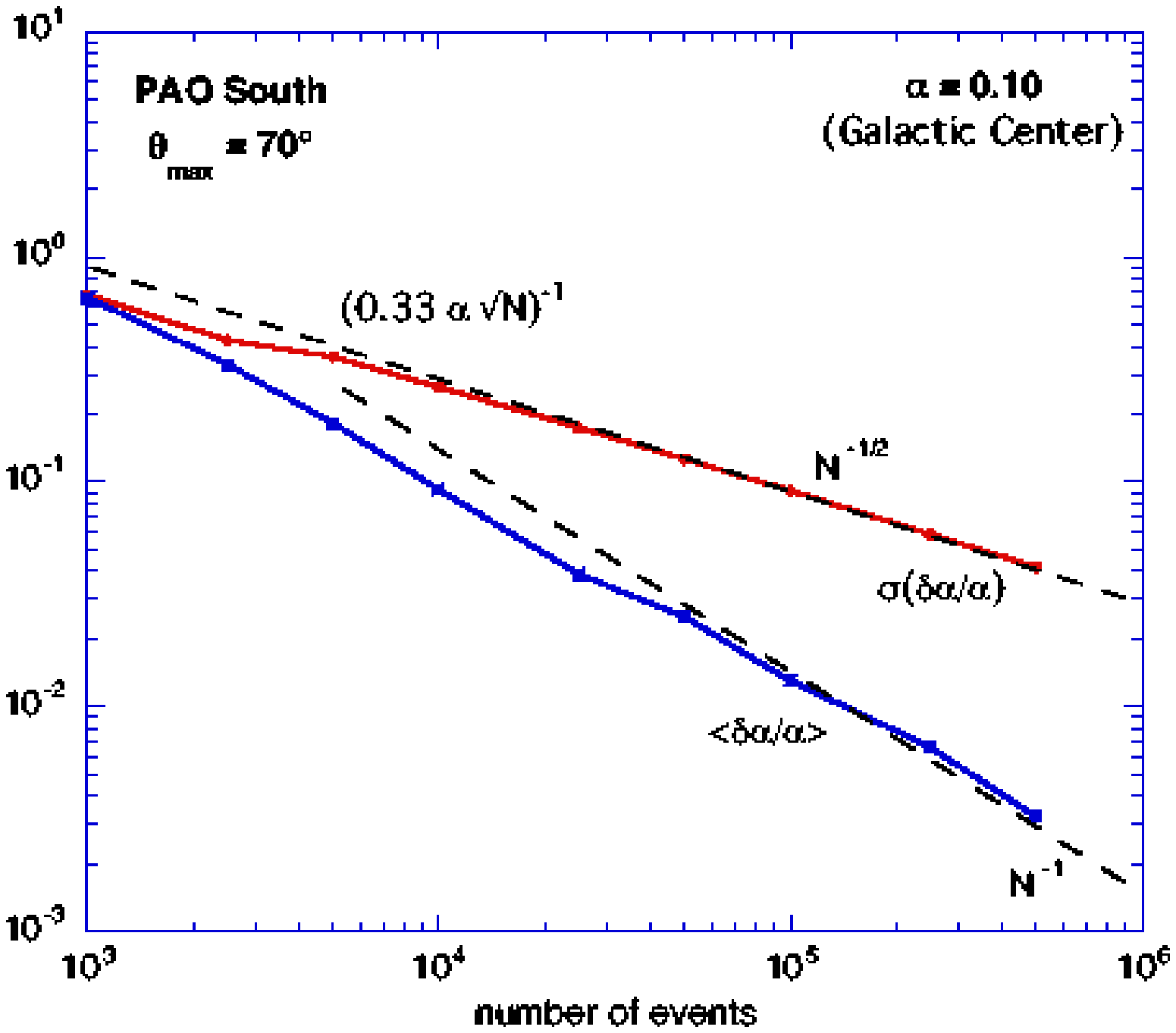}
\caption{Left: distributions of the relative error on the reconstructed dipole amplitude for various sizes of the data set, as indicated, in the case of the PAO detector with Southern site only, and a maximum zenith angle of $70^\circ$. Right: bias and dispersion of the reconstructed amplitude for a dipole with $\alpha = 0.10$ pointing towards the Galactic centre (GC), as a function of the number of events.}
\label{fig:deltaAlpha}
\end{figure*}

The power of Eqs.~(\ref{eq:alphaDxDyDz}) for a dipole reconstruction can be evaluated by generating random samples of events, and building the sums $S_{0}$ and $\mathbf{S}$ in Eq.~(\ref{eq:discreteSumsOverEventsFullSky}) to reconstruct the underlying dipole blindly (cf. Sommers, 2001).
We thus generated artificial data sets of various sizes, according to the anisotropic flux given in Eq.~(\ref{eq:CRDistribWithDipole}), taking also into account the relative exposure of the different parts of the sky as observed by any particular experiment of our choice. This comes down to drawing randomly $N$ directions over the sky, i.e. $N$ pairs of angles $(\theta_{i},\varphi_{i})$ representing the arrival direction of the cosmic-rays, with a probability $\mathcal{P}(\mathbf{u}) = \Phi(\mathbf{u})\times \mathcal{E}(\mathbf{u})$, where $\Phi(\mathbf{u})$ is given by Eq.~(\ref{eq:CRDistribWithDipole}) and $\mathcal{E}(\mathbf{u})$ depends on the experiment considered.

As an illustration, we consider the case of the Pierre Auger Observatory (PAO), either with the Southern site only or with both the Southern and Northern sites. We first calculate the corresponding exposure function for CRs at sufficiently high energy (so that the detection efficiency is saturated, i.e. independent of the CR arrival direction). The effective detection area is thus simply the ground area multiplied by $\cos\theta_{\mathrm{z}}$, where $\theta_{\mathrm{z}}$ is the local zenith angle. The exposure of a given point in the ``equatorial sky'' (declination, $\delta$, and right ascension, RA) is then obtained by integrating over the path of that point in the ``local sky'' over the detector, which depends on its latitude on Earth. The Southern site is located at $\lambda \simeq -35.2^{\circ}$, while the Northern site is not chosen yet, but should be at a similar (positive) latitude. The result is shown in Fig.~\ref{fig:expo}, for different assumptions concerning the maximum zenith angle up to which the detector is assumed to be working, namely $\theta_{\mathrm{z,max}} = 60^\circ$ and $80^\circ$. The PAO is expected to be working in standard reconstruction mode up to $\sim 70^\circ$, which is the value which we adopt for $\theta_{\mathrm{z,max}}$ in the calculations below (Auger Collaboration, 2004).

\begin{figure*}
\includegraphics[height=7.2cm]{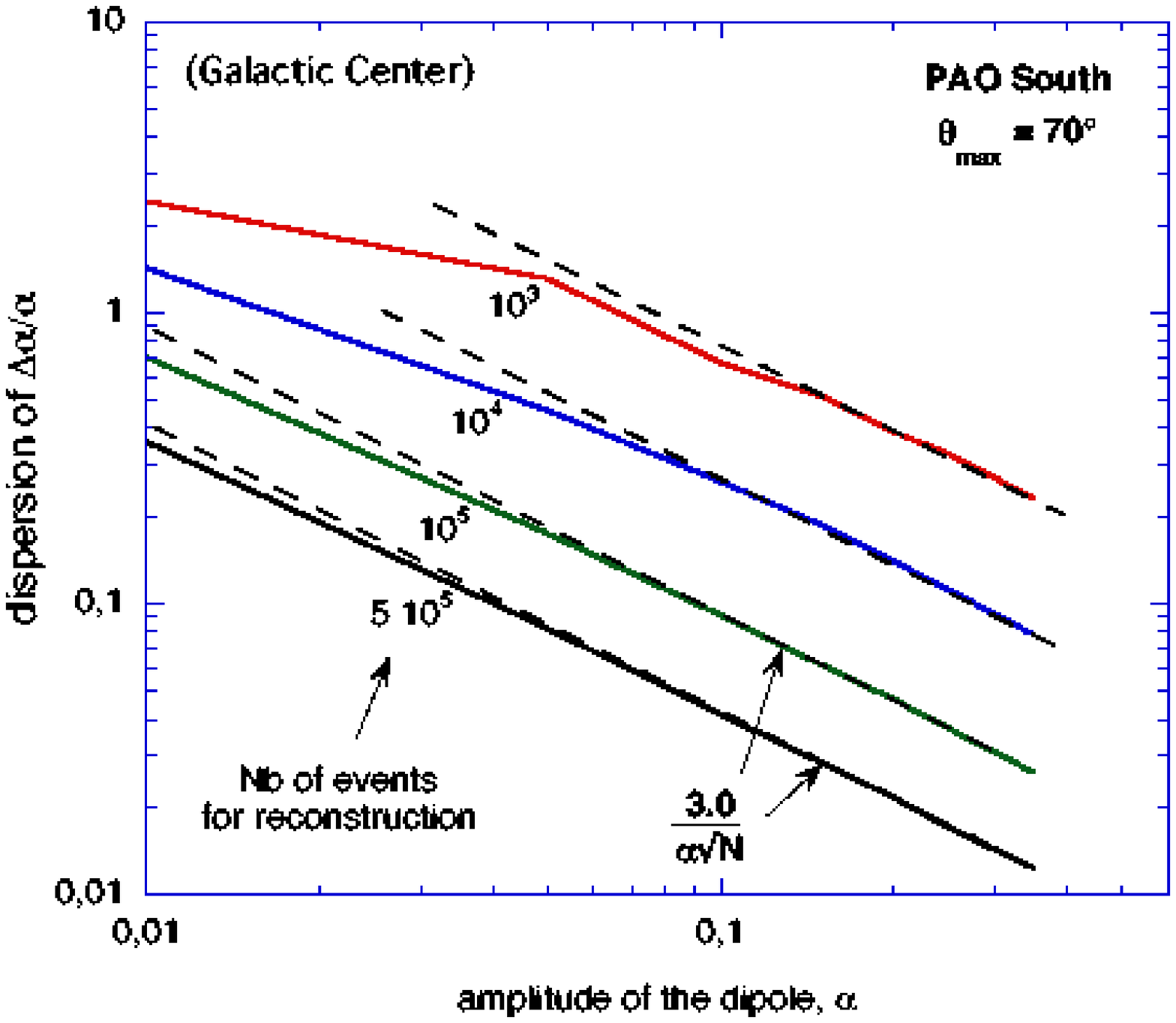}
\hfill
\includegraphics[height=7.2cm]{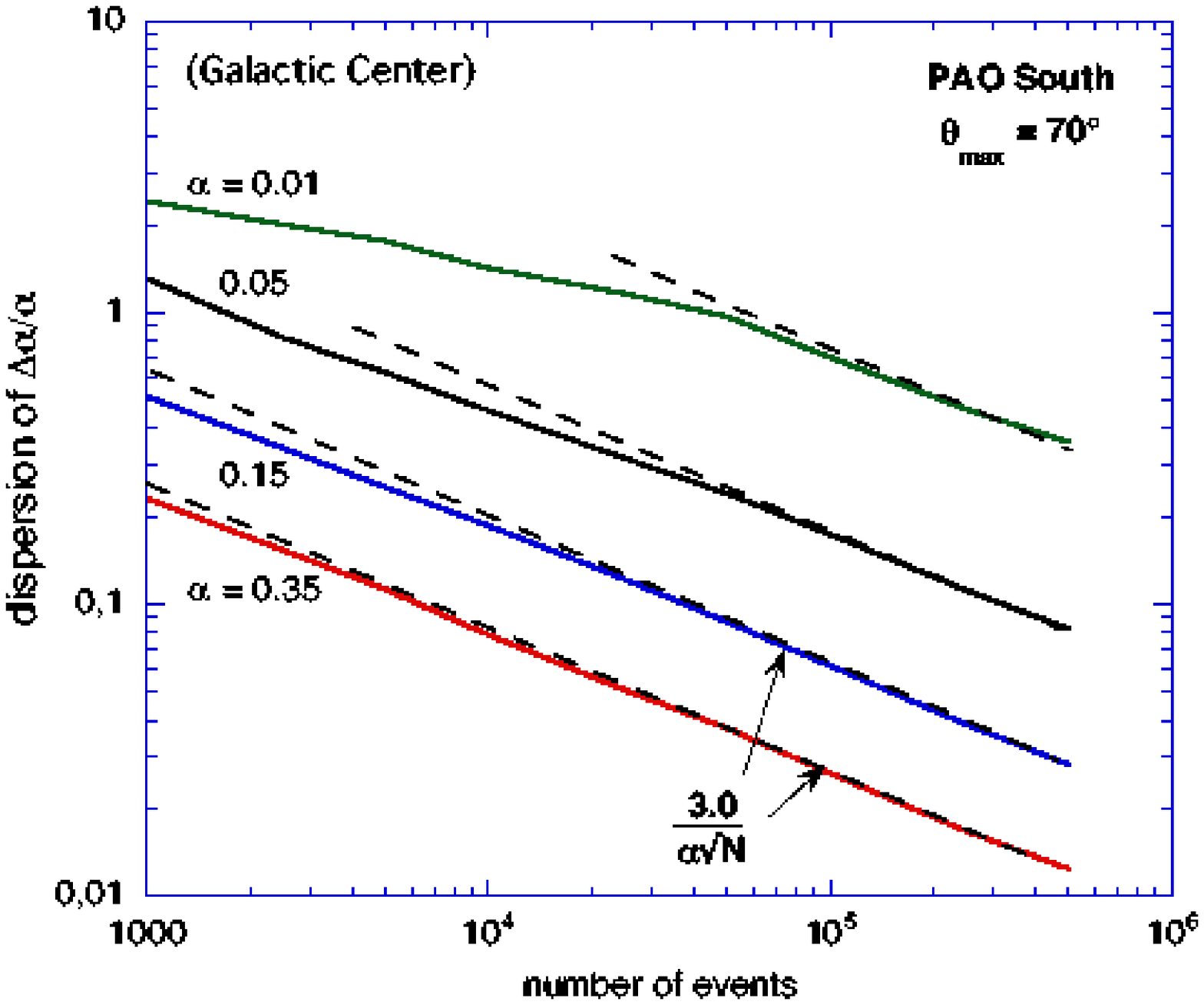}
\caption{Variation of the dispersion of $\delta\alpha/\alpha$, as a function of $\alpha$, for various sizes of the data set (left) and as a function of $N$, for various values of $\alpha$. The fit given by formula~(\ref{eq:power}) is also drawn.}
\label{fig:sigmaAlpha}
\end{figure*}

We then estimate the dipole reconstruction accuracy for various values of the dipole amplitude, $\alpha$, and direction, ($\theta_{\mathrm{d}}$,$\varphi_{\mathrm{d}}$), and for various sizes of the data set, $N$. In each case, we simulate a large number of data sets (namely $10^4$), and compare the reconstructed dipole parameters with the input values.

\subsection{Reconstruction of the dipole amplitude and direction}

The distribution of the reconstructed dipole amplitudes are shown on Fig.~\ref{fig:deltaAlpha}a for various values of $N$, in the case of a dipole oriented towards the GC with an amplitude of 10\% ($\alpha = 0.1$). The plotted quantity is the relative error on the reconstructed amplitude, namely $\delta\alpha/\alpha = (\alpha_{\mathrm{true}} - \alpha_{\mathrm{rec}})/\alpha_{\mathrm{true}}$. Its distribution is seen to be very close to Gaussian, even for relatively small data sets. The bias and dispersion both decrease when the data set gets larger. In Fig.~\ref{fig:deltaAlpha}b, this decrease is shown to follow respectively a $N^{-1}$ law and a $N^{-1/2}$ law at large $N$, as expected from statistics. Note also that the bias is always smaller than the dispersion.

On Fig.~\ref{fig:sigmaAlpha}, we show the dispersion of $\delta\alpha/\alpha$ as a function of $\alpha$, for various sizes of the data set, and as a function of $N$, for various values of $\alpha$. As can be seen, it obeys a general law $\sigma(\delta\alpha/\alpha) \propto (\alpha\sqrt{N})^{-1}$, which allows us to write the significance of a given measurement (number of sigmas, as in Eq.~\ref{eq:nSigmaR1h}):
\begin{equation}
n_{\sigma} = [\sigma(\delta\alpha/\alpha)]^{-1} = K_{\alpha}\,\alpha\,\sqrt{N},
\label{eq:power}
\end{equation}
where we have introduced the so-called \emph{reconstruction power}, $K_{\alpha}$, of the detector. This parameter fully characterises the ability of the detector under consideration to measure the amplitude of an underlying dipole anisotropy in the CR flux. In the case of a non uniform sky coverage, the reconstruction power also depends on the orientation of the dipole, as further discussed in the next subsection. For instance, in the case of a dipole oriented in the GC direction, observed by the Southern site of the PAO, the reconstruction power is $K_{\alpha} = 0.33$, as shown by the fits on Fig.~\ref{fig:sigmaAlpha}.

\begin{figure*}[t]
\includegraphics[height=7.2cm]{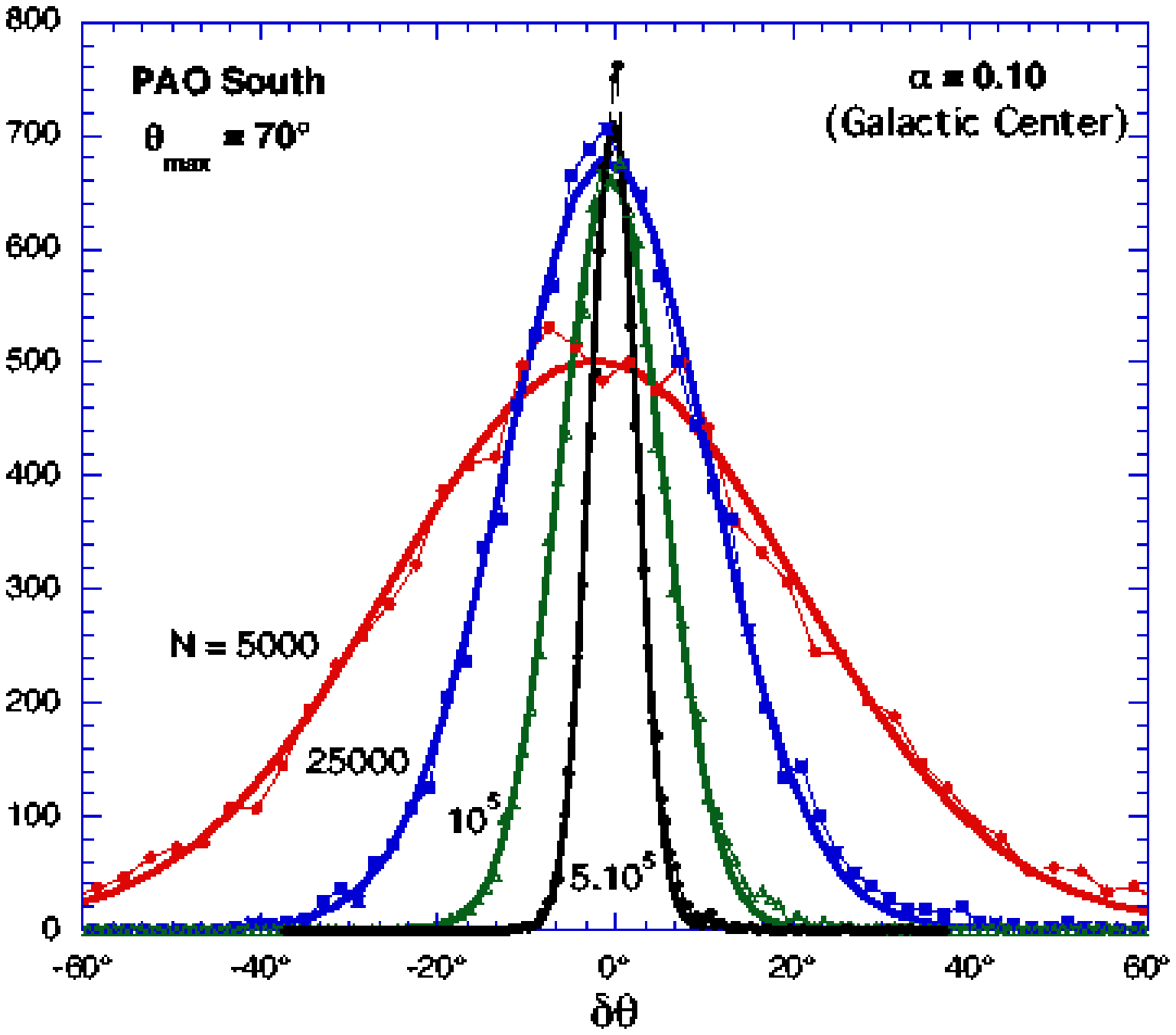}
\hfill
\includegraphics[height=7.2cm]{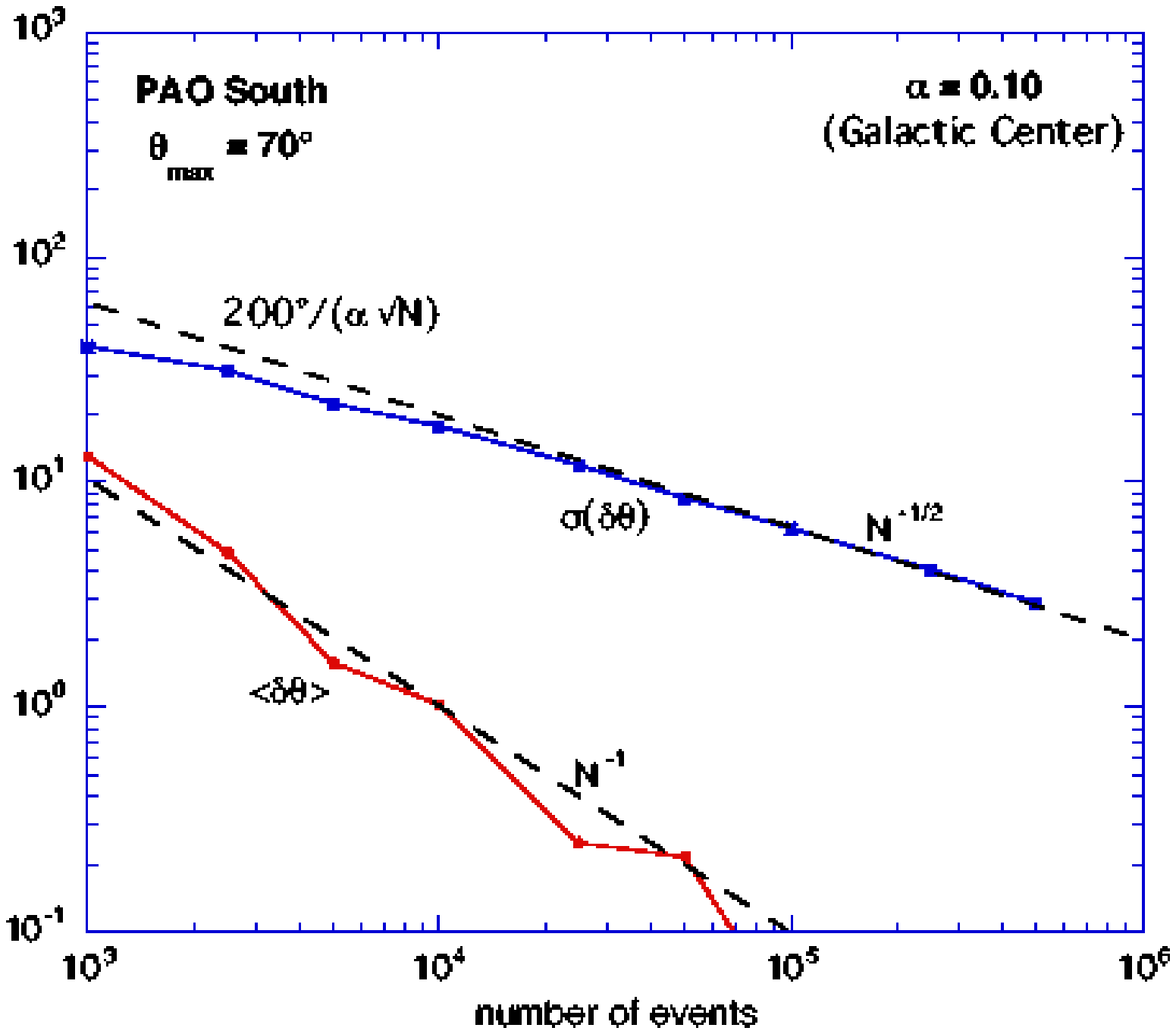}
\caption{Distributions of the relative error on the reconstructed dipole direction (left), and evolution of the corresponding bias and dispersion as a function of the number of events (right), for a dipole of amplitude $\alpha = 0.1$ oriented towards the GC, observed with the Southern site of  PAO detector with $\theta_{\mathrm{max}} = 70^\circ$.}
\label{fig:deltaTheta}
\end{figure*}

Likewise, we show on Fig.~\ref{fig:deltaTheta}a the distribution of the error on the reconstructed dipole declination, $\delta = \pi/2 - \theta$, for  various sizes of the data set, and in Fig.~\ref{fig:deltaTheta}b, the evolution of the bias and dispersion in $\delta$ (similar results are obtained with the other angular dimension, namely the right ascension of the dipole vector). A similar law is obtained for the reconstruction of the declination as for the amplitude:
\begin{equation}
\sigma_{\mathrm{dec}} = \frac{1}{K_{\mathrm{dec}}\,\alpha\,\sqrt{N}}.
\label{eq:KTheta}
\end{equation}
In the case of the Southern site of the PAO, with a dipole in the GC direction, one finds an angular power of $K_{\mathrm{dec}} \simeq 0.28\,\mathrm{rad}^{-1}$, or if one prefers $K_{\mathrm{dec}}^{-1} \simeq 200^\circ$. Likewise, for the reconstruction of the RA, one finds $K_{\mathrm{RA}} \simeq 0.38\,\mathrm{rad}^{-1}$, i.e. $K_{\mathrm{RA}}^{-1} \simeq 150^\circ$. For instance, if $\alpha = 0.05$ (as could be suggested by the AGASA data for CRs at $\sim 10^{18}$~eV; Hayashida et al. 1999) and for a data set of $10^5$ events, the angular reconstruction accuracy is about $12^\circ$ for $\delta$ and $9.4^\circ$ for RA.

\subsection{Comparison with the standard Rayleigh analysis}

In order to compare the power of our method with that of the standard analysis in right ascension, we first need to examine the link between $r_{\mathrm{1h}}$ and the dipole amplitude, $\alpha$. Indeed, while the significance of an anisotropy measurement has the same form in both cases, Eqs.~(\ref{eq:nSigmaR1h}) and~(\ref{eq:power}), it should be kept in mind that the first harmonic amplitude in right ascension has no clear physical meaning. It would thus be misleading to interpret Eq.~(\ref{eq:nSigmaR1h}) as demonstrating that the reconstruction power of the standard Rayleigh method is $1/\sqrt{2}\sim 0.707$. To compare the reconstruction power of the two methods, we need to evaluate the significance of the anisotropy measurement \emph{for the same value of the dipole amplitude}. Therefore, we need to relate $r_{\mathrm{1h}}$ and $\alpha$.

In the standard Rayleigh method, the basic quantities computed from the data are $a$ and $b$, given by Eq.~(\ref{eq:aAndB}). These are in fact discrete versions of continuous integrals which can be identified, according to the relations~(\ref{eq:integralsToSums}) and~(\ref{eq:nbOfEventsInPixelIJ}), as:
\begin{equation}
a= \frac{2}{N}\sum_{i=1}^N \cos\varphi_{i} = \frac{2}{N} \int \Phi(\mathbf{u}) \varepsilon(\mathbf{u}) \cos\varphi\d\Omega,
\label{eq:aContinuous}
\end{equation}
and
\begin{equation}
b= \frac{2}{N}\sum_{i=1}^N \sin\varphi_{i} = \frac{2}{N} \int \Phi(\mathbf{u}) \varepsilon(\mathbf{u}) \sin\varphi\d\Omega,
\label{eq:bContinuous}
\end{equation}
where $N$ is given by (see Eq.~\ref{eq:nbOfEventsInPixelIJ}):
\begin{equation}
N= \int \Phi(\mathbf{u}) \varepsilon(\mathbf{u}) \d\Omega.
\label{eq:N}
\end{equation}

The first harmonic amplitude, $r_{\mathrm{1h}} = \sqrt{a^2 + b^2}$, is thus obtained by direct integration of the above formul\ae, over the part of the sky where the exposure, $\mathcal{E}(\theta)$ (assumed independent of RA), is non zero. For a dipole orientation $\theta_{\mathrm{d}}$ (in equatorial spherical coordinates), one obtains:
\begin{equation}
r_{1h} = \left |\frac {c_{3}\, \alpha \sin \theta_{\mathrm{d}}}{c_{1} + c_{2} \,\alpha \cos \theta_{\mathrm{d}}}\right |
\label{eq:r1hVSalpha}
\end{equation} 
where $c_{1}$, $c_{2}$ and $c_{3}$ are numerical constants which only depend on the exposure function of the detector under consideration, and are given by:
\begin{equation}
\begin{split}
c_{1} &= \int _{\theta_{min}}^{\theta_{max}} \mathcal{E}(\theta) \sin \theta \,\, \d\theta \\
c_{2} &= \int _{\theta_{min}}^{\theta_{max}} \mathcal{E}(\theta) \sin \theta \cos \theta \,\,\d\theta \\
c_{3} &= \int _{\theta_{min}}^{\theta_{max}} \mathcal{E}(\theta) \sin ^2\theta\,\, \d\theta
\end{split}
\label{eq:integralesE}
\end{equation} 

In the case of the PAO detector with the Southern site only, we find: $c_{1} = 0.77$, $c_{2} = -0.32$ and $c_{3} = 0.60$. For the full Auger detector (both sites), we have: $c_{1} = 1.4$, $c_{2} = 0.028$ and $c_{3} = 1.1$. For a full-sky uniform exposure, the limiting values would be $c_{1} = 2$, $c_{2} = 0$ and $c_{3} = \pi/2$.

As can be seen, the relation between $r_{\mathrm{1h}}$ and $\alpha$ (the true, physical dipole amplitude) depends on the dipole declination, $\delta_{\mathrm{d}} = \frac{\pi}{2} - \theta_{\mathrm{d}}$, which is precisely \emph{not} reconstructed by the method. It also appears that, as expected, $r_{1h}$ vanishes for $\delta_{d}=-\pi/2$ and $\delta_{d}=\pi/2$, which indicates that the right ascension analysis is totally inadequate to study the anisotropy of the data set in such a case. This is obviously because when the dipole vector is oriented along the rotation axis of the Earth, the flux modulation is only in declination. The efficiency of the dipole reconstruction with such a method is thus expected to be largest when the dipole vector is in the equatorial plane.

We can now compare the reconstruction powers of the standard Rayleigh method and our 3D analysis for different dipole orientations. To this purpose, we simply rewrite Eq.~(\ref{eq:nSigmaR1h}) as a function of $\alpha$:
\begin{equation}
n_{\sigma} = \frac{1}{\sqrt{2}}\,r_{\mathrm{1h}}\,\sqrt{N}\, \equiv \,K_{\mathrm{1h}}\,\alpha\,\sqrt{N},
\label{eq:nSigmaR1hBis}
\end{equation}
with
\begin{equation}
K_{\mathrm{1h}} = \frac{1}{\sqrt{2}}\,\frac{r_{\mathrm{1h}}}{\alpha} = \frac{1}{\sqrt{2}}\,\left |\frac {c_{3}\, \sin \theta_{\mathrm{d}}}{c_{1} + c_{2} \,\alpha \cos \theta_{\mathrm{d}}}\right |.
\label{eq:KR1h}
\end{equation}

\begin{figure}[t]
\begin{center}
\epsfig{file=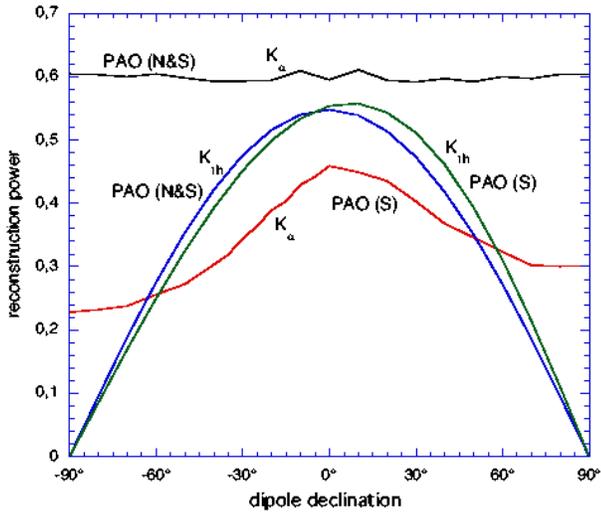,width=8cm}
\end{center}
\caption{Comparison of the dipole amplitude reconstruction powers as a function of the dipole declination, in the case of the PAO with one (S) or two sites (N\&S). The power of the Rayleigh method, $K_{\mathrm{1h}}$, is given by Eq.~(\ref{eq:KR1h}).}
\label{fig:reconstructionPower}
\end{figure}

Figure~\ref{fig:reconstructionPower} shows the dipole amplitude reconstruction power both with our method and with the standard one, as a function of the dipole declination, in the case of the PAO with one or two sites (note that only the Southern site is funded yet, and close to completion, while the Northern site in the selection process). As can be seen, in the case of a partial sky coverage (Southern site only), the first harmonic amplitude method is more powerful for dipole declinations between $-60^\circ$ and $+60^\circ$. However, it should be remembered that only one angle is reconstructed in this case, namely the right ascension, and that the inferred $r_{\mathrm{1h}}$ is not the dipole amplitude (which \emph{cannot} be derived). For other dipole orientations, our method is more powerful (essentially because almost no anisotropy remains along RA coordinate), and reconstructs entirely the dipole direction, together with its true amplitude.

The reconstruction power, $K_{\alpha}$, still gets smaller when the dipole vector is aligned with the Earth rotation axis, but it does not drop to zero, as the 3D analysis also makes use of the declination information in the CR arrival directions. The amplitude of the declination dependence of $K_{\alpha}$ is limited to a factor of $\sim 2$.

In the case when both PAO sites are available, our reconstruction power is always larger than with the Rayleigh method, and essentially independent of the dipole orientation. With such a full-sky experiment, the reconstructed method presented here reduces exactly to that of Sommers (2001), and it is therefore natural that we find essentially the same value for $K_{\alpha}$. The value quoted by Sommers (2001) is 0.65, instead of $\sim 0.61$ in our case, but his value was obtained with a smaller data set, and as can be seen on Figs.~\ref{fig:deltaAlpha}b and \ref{fig:sigmaAlpha}b, the effective reconstruction power for smaller data sets is always slightly higher than the asymptotic value obtained at large $N$ (which is the one we plot on Fig.~\ref{fig:reconstructionPower}). Therefore, we can claim excellent agreement with previous work when a comparison is possible (i.e. for full sky exposure experiments).

From the experimental point of view, a larger reconstruction power implies that a given dipole amplitude will be detected earlier, i.e. with a smaller number of events. Quantitatively, adding the Northern site to the Pierre Auger Observatory will allow one to detect a dipole anisotropy with a power between $1.3$ and $2.7$ times larger, or a factor of 2 on average (for an unknown dipole orientation). According to Eq.~(\ref{eq:power}), this means that its identification will require between $1.7$ and $7.3$ times less events, or a factor of 4 on average. Considering that the Northern site will also double the PAO acceptance, and thus the rate of events detection, one can expect on average that the complete PAO detector will be able to measure a dipole anisotropy (at a given significance level) about 8 times quicker than just its Southern site (or between 3.4 and 14.6 depending on the dipole orientation).

\begin{figure*}[t]
\includegraphics[height=7.2cm]{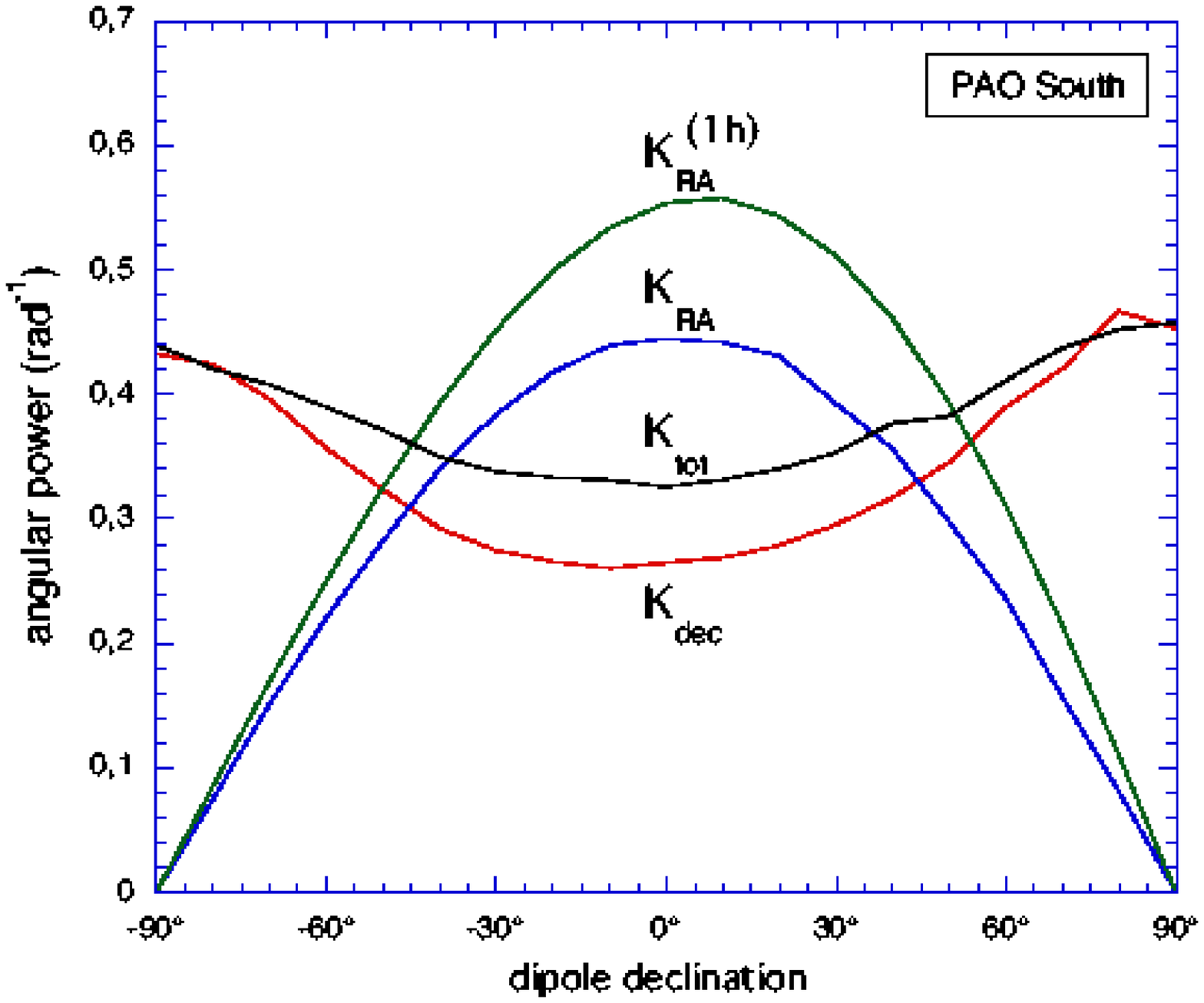}
\hfill
\includegraphics[height=7.2cm]{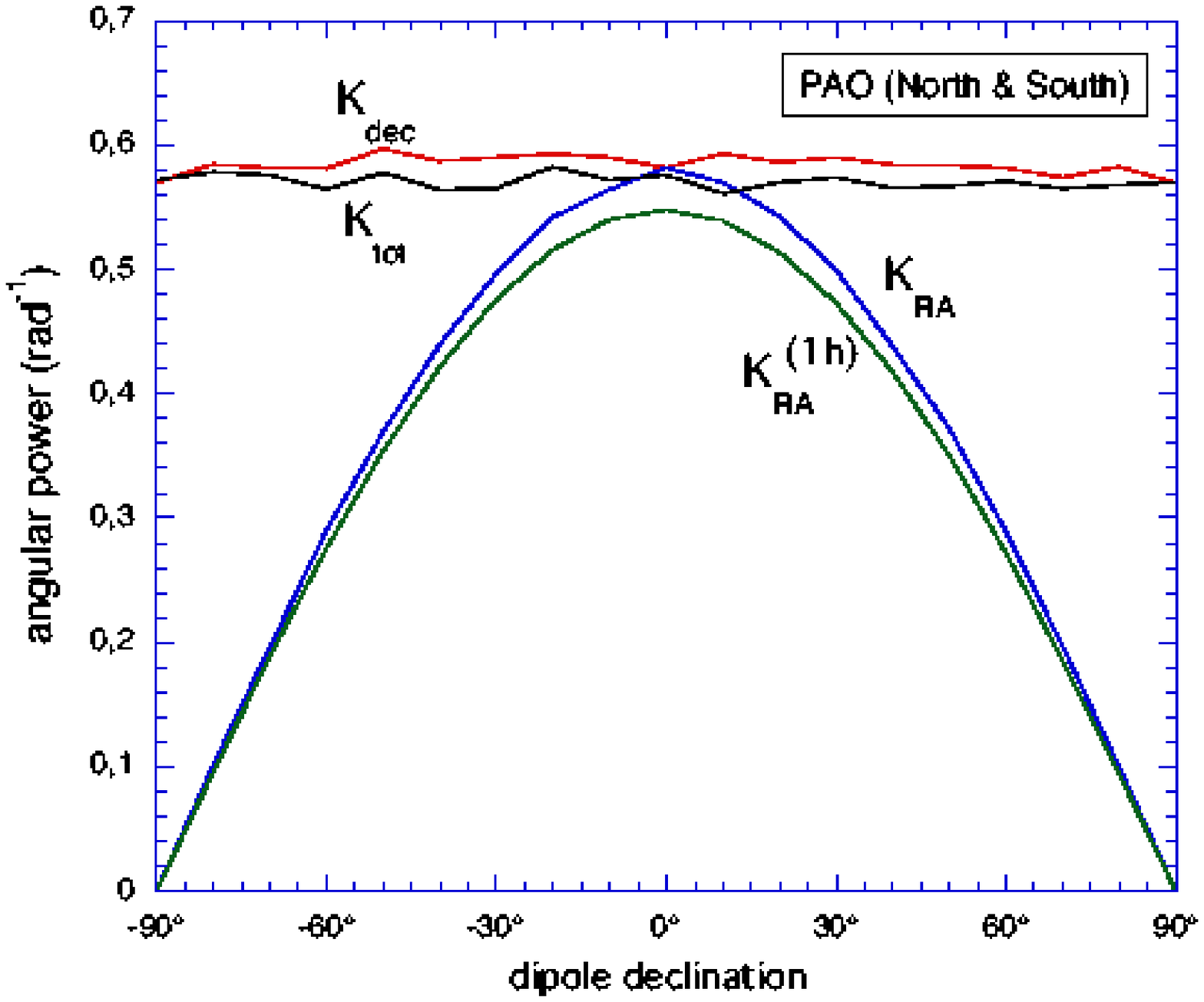}
\caption{Angular reconstruction powers of the PAO with one (left) or two sites (right), as a function of the dipole declination. The labels refer to the total angular accuracy, and separately to its declination and RA parts, compared to the RA accuracy of the first harmonic method.}
\label{fig:angularAccuracy}
\end{figure*}

Concerning the angular accuracy, the reconstruction power is plotted in Fig.~\ref{fig:angularAccuracy}a  and \ref{fig:angularAccuracy}b as a function of the dipole declination, for either one or two PAO sites. In both cases, we show the angular power for the reconstruction of the dipole declination, $K_{\mathrm{dec}}$, and right ascension, $K_{\mathrm{RA}}$, together with the total angular power in 3D, $K_{\mathrm{tot}}$. The latter is obtained from the 3D angle, $\gamma$, between the true and the reconstructed dipole directions, given by $\cos\gamma = \cos\theta_{\mathrm{d}}\cos\theta_{\mathrm{rec}} + \sin\theta_{\mathrm{d}}\sin\theta_{\mathrm{rec}}\cos(\varphi_{\mathrm{d}} - \varphi_{\mathrm{rec}})$. For comparison, we also plotted the pseudo angular power of the standard Rayleigh method (relative to the RA projection of the dipole direction), which is obtained from the angular accuracy on the right ascension recalled in Sect.~\ref{sec:RayleighMethod}, $\sigma_{\psi} = \sqrt{2/Nr_{\mathrm{1h}}^2}$:
\begin{equation}
\begin{split}
K_{\mathrm{RA}}(\mathrm{1h}) &= \frac{\sigma_{\psi}^{-1}}{\alpha\sqrt{N}} = \frac{1}{\sqrt{2}}\,\frac{r_{\mathrm{1h}}}{\alpha} \\
&= \frac{1}{\sqrt{2}}\,\left |\frac {c_{3}\, \sin \theta_{\mathrm{d}}}{c_{1} + c_{2} \,\alpha \cos \theta_{\mathrm{d}}}\right |\,\,\mathrm{rad}^{-1}.
\end{split}
\label{eq:KRA1h}
\end{equation}

Note that it has the same formal expression as $K_{\mathrm{1h}}$, in Eq.~(\ref{eq:KR1h}), but it is essentially a different quantity, expressed in $\mathrm{rad}^{-1}$, while the amplitude reconstruction power is dimensionless.

In the case of the Southern site alone, we first see that the reconstruction of the dipole right ascension is always more accurate with the first harmonic amplitude method, as the full statistics is used to derive this parameter alone. On the other hand, only the 3D method can reconstruct the dipole declination and it is found that the accuracy on this angle is better than that on RA for dipole orientations with $\delta_{\mathrm{d}} \la -45^\circ$ or $\delta_{\mathrm{d}} \ga 45^\circ$. As for the dipole amplitude reconstruction, the CR flux anisotropy is poorly analysed by the 2D Rayleigh method when the dipole vector is at a small angle from the Earth rotation axis, since only a very small modulation then remains on the RA coordinate.

This transfer of the main flux modulation from one angular coordinate to the other is also responsible for the anti-correlation, observed on Fig.~\ref{fig:angularAccuracy}, between the reconstruction powers for RA and $\delta$. The total angular power, given by $K_{\mathrm{tot}}$, shows a smaller dependence on the dipole declination, of the order of $\pm 15$\% in the case of the PAO Southern site, with a maximum when the dipole is oriented towards one of the poles.

With the two PAO sites, the angular reconstruction power is roughly independent of the dipole declination (as the exposure is almost flat over the whole sky), and always larger with our 3D method than with the standard RA analysis. Even when the latter is at its maximum, namely when the dipole vector is in the equatorial plane, the 3D reconstruction -- i.e. the reconstruction of two angles instead of one -- is not less accurate than the standard 2D Rayleigh reconstruction. Quantitatively, in the case of a dipole amplitude of 5\% and a data set of $10^5$ events, the PAO with two sites would give an accuracy of $\simeq 6.4^\circ$, instead of $\simeq 11^\circ$ with only one site.

\section{Conclusion}

We have presented a simple numerical method allowing one to reconstruct the parameters of a dipolar modulation of an otherwise isotropic CR angular distribution. In contrast with the standard Rayleigh analysis making use of the distribution of CRs in right ascension only, our method enables a full 3D reconstruction, providing a direct determination of the underlying dipole right ascension and declination, as well as of its true amplitude. We also calculated the relation between the \emph{first harmonic amplitude in right ascension}, $r_{\mathrm{1h}}$, as obtained from the 2D Rayleigh method, and the \emph{true dipole amplitude}, $\alpha$, as a function of the underlying dipole orientation. This relation strongly depends on the dipole declination, which cannot be reconstructed by the standard method.

The method presented here is essentially a generalisation of that of Sommers (2001), which can now be applied in situations where the experimental data do not cover the entire sky. In particular, it can be applied with data sets obtained from Earth-based experiments detecting CRs arriving from regions bounded by lines of equal declination in the equatorial sky.

In order to quantify the accuracy of the method, we introduced the concept of \emph{reconstruction power}, for the dipole amplitude as well as for its direction (in 3D space). The significance of an anisotropy measurement can be expressed as $n_{\sigma} = K_{\alpha}\alpha\sqrt{N}$, where $\alpha$ is the dipole amplitude, $N$ is the number of events detected, and $K_{\alpha}$ is the amplitude reconstruction power, which depends on the experiment under consideration, through its \emph{exposure function}, $\mathcal{E}(\mathbf{u})$, giving the achieved relative exposure of the different regions of the sky. We have shown that the reconstruction power depends on the orientation of the dipole underlying the CR angular distribution, but does not drop to zero when the dipole is aligned with the Earth rotation axis, as with the standard 2D Rayleigh method.

We also investigated the angular accuracy, and showed that it can be expressed as $\sigma_{\mathrm{tot}} = K_{\mathrm{tot}}\alpha\sqrt{N}$, where the angular reconstruction power, $K_{\mathrm{tot}}$, also depends on the dipole declination, but to a smaller extent. In particular, the accuracy of the method presented here can be higher than that of the standard Rayleigh method, even though it uses the same data set to reconstruct one more angular parameter, as well as the true (physical) dipole amplitude. Therefore, we believe that this is a powerful method to be widely used for angular distribution analysis, in particular in CR physics.

An obvious flaw of the method, however, is that it \emph{assumes} a purely dipolar modulation of an isotropic flux (without any restriction on the dipole amplitude). As was discussed in the Introduction, higher order spherical harmonics can be present in the underlying angular distribution, which would then partly spoil the reconstruction of the dipolar term. In particular, a purely quadrupolar distribution could be mistakenly reconstructed as a dipole. However, it may first be argued that there is a wide range of situations where the dipolar term should be dominant in the CR angular distribution, as one might expect it to be the first one to be detectable when perfect isotropy is lost (unless particular symmetry conditions apply). It should be noted also that the situation is the same with the standard Rayleigh method, which not only cannot indicate the nature of the detected anisotropy (i.e. whether it is a dipole or it involves higher order terms), but cannot even provide the dipole amplitude when the anisotropy is a purely dipolar modulation of an isotropic flux. Nevertheless, even if the dominant anisotropy is not dipolar, both the first harmonic amplitude analysis in right ascension and our 3D method remain useful to determine the significance of the observed departure from isotropy. With the formalism introduced in this paper, if a ``dipole'' amplitude $\alpha$ is reconstructed from a data set containing $N$ events, the CR angular distribution can be safely stated to be anisotropic with a statistical significance of $n_{\sigma} = K_{\alpha} \alpha\sqrt{N}$. Note also that higher order spherical harmonics can be investigated through a systematic study of the CR angular power spectrum, as reported recently by Deligny et al. (2004).

Finally, as an illustration of our method, we computed the angular and amplitude reconstruction powers of the Pierre Auger Observatory, comparing the values obtained with the Southern site alone or with both Southern and Northern sites. We found that the Northern site will lead to a considerable increase of power, by a factor of $\sim 2$ on average (depending on the underlying dipole orientation) for the reconstruction of both the amplitude and the direction (in 3D space). This implies that our method will assign a given statistical significance to an anisotropy measurement with a data set $\sim 4$ times smaller if two sites are available instead of one. A given measurement will thus be reached 8 times quicker with two sites (taking into account the corresponding increase in the acceptance) or 4 times quicker than if the Southern site has its area doubled, rather than duplicated in the Northern hemisphere. Conversely, for a given size of the overall data set, an anisotropy will be measured with typically twice as many sigmas and the angular resolution will be twice as more precise with a full sky detector than with a partial sky experiment such as the PAO Southern site. This may be considered as another important argument in favour of the development of the PAO Northern site.

\begin{acknowledgements}
We thank the Auger Collaboration for stimulating discussions about CR anisotropy measurements, and for reading and commenting the paper.
\end{acknowledgements}

\end{document}